\documentclass[12pt]{article}
\usepackage{amsmath,amssymb}
\usepackage{graphicx,color}
\numberwithin{equation}{section}
\usepackage{cite}
\usepackage{bm}
\usepackage{dcolumn}  
\newcommand{\beq}{\begin{equation}}
\newcommand{\eeq}{\end{equation}}
\newcommand{\beqa}{\begin{eqnarray}}
\newcommand{\eeqa}{\end{eqnarray}}
\newcommand{\beqar}{\begin{eqnarray*}}
\newcommand{\eeqar}{\end{eqnarray*}}
\newcommand{\al}{\alpha}

\newcommand{\ka}{\kappa}

\renewcommand{\l}{\lambda}

\newcommand{\ie}{{\it i.e.,}\ }
\newcommand{\labell}[1]{\label{#1}} 
\newcommand{\reef}[1]{(\ref{#1})}
\newcommand\prt{\partial}

\begin{document}
\begin{titlepage}
\hfill
\vbox{
    \halign{#\hfil         \cr
           hep-th/yymmnnn\cr
           IPM/P-2007/023 \cr
           } 
      }  
\vspace*{20mm}
\begin{center}
{\Large {\bf On Attractor Mechanism and Entropy Function for Non-extremal Black Holes/Branes}\\ }

\vspace*{15mm} 
\vspace*{1mm} 
{\large Mohammad R. Garousi and Ahmad Ghodsi}
\vspace*{1cm}

{ {Department of Physics, Ferdowsi University, P.O. Box 1436, Mashhad, Iran}}\\
{ and}\\
{ Institute for Studies in Theoretical Physics and Mathematics (IPM)\\
P.O. Box 19395-5531, Tehran, Iran}\\

\vskip 0.6 cm

E-mail: {{ garousi@ipm.ir, ahmad@ipm.ir}} \\
\vspace*{1cm}
\end{center}

\begin{abstract} 
We examine in details the entropy function formalism for non-extremal $D3$, $M2$, and 
$M5$-branes that their throat approximation is given by Schwarzschild 
black hole in $AdS_{p+2}\times S^{D-(p+2)}$. 
We show that even though there is no attractor mechanism in the non-extremal 
black holes/branes, 
the entropy function formalism does work and the 
entropy is given by the entropy function at its saddle point. 
\end{abstract} 

\end{titlepage} 

\section{Introduction} \label{intro} 
The black hole attractor mechanism has been an active subject over the past 
few years in string theory. This is originated  from the observation that there 
is a connection between the partition function of four-dimensional BPS black 
holes and partition function of topological strings \cite{Ooguri:2004zv}. 

The attractor mechanism states that in the extremal black hole backgrounds 
the moduli scalar fields at horizon are determined by the charge of black 
hole and are independent of  their asymptotic values. 
One may study the attractor mechanism by finding the effective 
potential for the moduli fields and examining the behavior of the effective 
potential at its extremum, \ie in order to have the attractor mechanism, 
the effective potential must have minimum in all directions. The entropy of black hole is 
then given by the value of the effective potential at its minimum. Using 
this, the entropy of some extremal black holes has been calculated in \cite{attractor}. 

Motivated by the attractor mechanism, it has been proposed  by A. Sen that 
the entropy of a specific class of extremal black holes in higher derivative 
gravity can be calculated using the entropy function formalism \cite{Sen}. 
According to this formalism, the entropy function for the black holes that 
their near horizon is $AdS_2\times S^{D-2}$ is defined by integrating the 
Lagrangian density over $S^{D-2}$ for a general $AdS_2\times S^{D-2}$ 
background characterized by the size of $AdS_2$ and $S^{D-2}$,  and taking 
the Legendre transform  of the resulting function with respect to the 
parameters labeling the electric fields. The result is  a function of   
moduli scalar fields as well as the size of $AdS_2$ and $S^{D-2}$. The 
values of moduli fields and the sizes are determined by extremizing 
the entropy function with respect to the moduli fields 
and the sizes. Moreover, the entropy is given by 
the value of the entropy function at the extremum\footnote{It is assumed 
that in the presence of higher derivative terms there is a solution whose near 
horizon geometry is $AdS_2\times S^{D-2}$. In the cases that the higher 
derivative corrections modify the solution such that the near horizon is 
not $AdS_2\times S^{D-2}$ anymore, one cannot use the entropy function formalism. 
In those cases one may use the Wald formula \cite{Wald} to calculated the 
entropy directly.}. Using this method the entropy of some extremal black holes have been found
in \cite{Sen},\cite{Ghodsi:2006cd},\cite{entfunc}.

For non-extremal black holes, one expects to have no   
attractor mechanism. 
An intuitional explanation of attractor mechanism
has been proposed in \cite{Kallosh:2006bt}. According to which the 
physical distance from an arbitrary point to the horizon is infinite for black holes
which have attractive horizon. 
While the physical distance is infinite 
for extremal black holes, it is finite for non-extremal cases.  
Alternatively, it has been shown in \cite{Goldstein:2005hq} that the values of 
the moduli fields at the horizon of non-extremal black holes depend on the 
asymptotic values of the scalar fields, hence, one expects to have no 
attractor mechanism for the non-extremal cases. 

It is natural to ask if the entropy function formalism works for a non-extremal 
black hole. 
We speculate that the entropy function formalism works if 
the background is some extension of $AdS$ at its near horizon. Moreover, for this background the entropy
function has saddle point at the near horizon.
In general, non-extremal black hole/brane solutions can be classified into 
three classes: 
1) Solutions with no moduli, 
2) Solutions with constant moduli, 
3) Solutions with constant moduli at the near horizon.
In this paper, we would like to consider the non-extremal
black-branes whose near horizons are Schwarzschild black hole in $AdS_{p+2}\times S^{D-(p+2)}$.     
For $p=3$, the solution is the non-extremal $D3$-brane with constant moduli. For $p=2,5$, the solutions are the non-extremal $M2$ and $M5$-branes with no moduli. We will 
discuss also the non-extremal black hole solutions 
with constant moduli at the near horizon which has been considered in \cite{Cai:2007ik}.

An outline of the paper is as follows. In section 2, we review the non-extremal solutions of 
IIB/M theory. In sections 3 to 5, using the entropy function formalism we derive the known 
results for the entropy of $D3$, $M2$ and $M5$-branes in terms of the temperature. We also 
show that in all cases the entropy is given by the entropy function at its saddle point. In 
section 6  we show that the higher derivative terms do not respect the symmetries of the 
solution at tree level and so the entropy function formalism does not work. Instead, we use 
the Wald formula directly to find the correction to the entropy. 
We conclude with a discussion of our results in the last section.

\section{Review of the non-extremal solutions}
In this section, we review the non-extremal solutions of IIB/M theory. The two-derivative 
effective action for IIB/M theory in Einstein frame is given by
\beqa 
S&=&\frac{1}{16\pi G_{D}}\int d^{D}x\,\sqrt{-g}\bigg\{
R-\frac{1}{2}g^{\mu\nu}\prt_{\mu}\phi\prt_{\nu}\phi-\frac{1}{2}
\sum\frac{1}{n!}F_{(n)}^2+\cdots\bigg\}\,,
\eeqa
where $D=10$ for IIB and $D=11$ for M theory. In above Lagrangian $\phi$ is the dilaton 
which appears only in IIB theory, and $F_{(n)}$ is the electric field 
strength where $n=1,3,5$ for 
IIB theory and $n=4,7$ for M theory. The $n=5$ field
strength tensor is self-dual, hence, it is not described by the
above simple action. It is sufficient to adopt the
above action for deriving the equations of motion, and impose the
self-duality by hand. Dots represent fermionic terms as well as NS-NS 3-form field
strength for IIB theory.

We are interested in non-extremal solutions whose near horizon are product space of $AdS$ with a 
sphere. $D3$, $M2$ and $M5$-branes have this property.
These solutions are given by the following (see e.g. \cite{Petersen:1999zh}):
\beqa
ds^2&=&H^{-\frac{d-2}{D-2}}\bigg(-f dt^2+\sum_{i=1}^{p}(dx^i)^2\bigg)
+H^{\frac{p+1}{D-2}}\bigg(f^{-1}dr^2+r^2(d\Omega_{d-1})^2\bigg)\,,\cr &&\cr
e^{\phi}&=&1\,\,,\qquad\qquad F_{ti_1\cdots i_pr}\,\,=\,\,\epsilon_{i_1\cdots
i_p}H^{-2}\frac{Q}{r^{d-1}}\,,\cr &&\cr
H&=&1+\bigg(\frac{h}{r}\bigg)^{d-2}\,\,,\qquad f=1-\bigg(\frac{r_0}{r}\bigg)^{d-2}\,,
\eeqa
where $D=(p+1)+d$ and $d$ is the number of dimensions transverse to the p-brane. Note that 
for $p=3$, the above field strength is only the electric part of the self-dual $F_{(5)}$. We 
will see shortly that in the entropy function formalism one needs to consider only this part 
of $F_{(5)}$. The relation between $h$ and $Q$ is 
\beqa
h^{2(d-2)}+h^{d-2}r_0^{d-2}&=&\frac{Q^2}{(d-2)^2}\,.\labell{relation}
\eeqa
For $r_0=0$ we obtain the extremal solution, depending only on a
single parameter, $Q$, related to the common mass and charge
density of the BPS p-branes. 

For $r_0\neq 0$ a horizon develops at $r=r_0$.  The near horizon geometry which 
is described by a throat 
can be found by using the throat approximation where $r\ll h$.
In this limit the relation \reef{relation} simplifies to
$h^{d-2}=Q/(d-2)$, and the non-extremal solution becomes 
\beqa
ds^2&=&\left(\frac{r}{h}\right)^{\frac{2(d-2)}{p+1}}\bigg\{
-\left[1-\left(\frac{r_0}{r}\right)^{d-2}\right]dt^2
+\sum_{i=1}^{p}(dx^i)^2\bigg\} +\left(\frac{h}{r}\right)^{2}
\left[1-\left(\frac{r_0}{r}\right)^{d-2}\right]^{-1}dr^2\cr &&\cr
&+&h^2(d\Omega_{d-1})^2\,,\cr &&\cr
e^{\phi}&=&1\,\,,\qquad F_{ti_1\cdots
i_pr}\,\,=\,\,(d-2)\epsilon_{i_1\cdots
i_p}\frac{r^{d-3}}{h^{d-2}}\,,
\eeqa 
where the geometry is the product of $S^{d-1}$ with the Schwarzschild black hole in $AdS_{D-d+1}$.


\section{Entropy function for non-extremal $D3$-branes}
Following \cite{Sen}, in order to find the entropy function for non-extremal 
$D3$-branes one can deform the near horizon geometry as  
\beqa
ds^2_{10}\!\!\!&=&\!\!\!v_1\left[\frac{r^2}{h^2}\left\{-\left(1-\left(\frac{r_0}{r}\right)^4\right)
dt^2 +\sum_{i=1}^{3}(dx^i)^2\right\}
+\frac{h^2}{r^2}\left(1-\left(\frac{r_0}{r}\right)^4\right)^{-1}
dr^2\right]
+v_2h^2(d\Omega_5)^2\!,\cr &&\cr
e^{\phi}\!\!\!&=&\!\!\!1\,\,,\qquad
F_{ti_1i_2i_3r}\,\,=\,\,4\epsilon_{i_1i_2i_3}\left(\frac{v_1}{v_2}\right)^{5/2}\frac{r^3}{h^4}\equiv
\epsilon_{i_1i_2i_3}e_1\,,\labell{solnxd3}
\eeqa 
where $v_1$ and $v_2$ are supposed to be constants. Note that we have considered 
only the electric part of the self-dual $F_{(5)}$. The function $f$ is define to be 
the integral of Lagrangian density over the horizon $H=S^3\times S^5$. The result of 
inserting the background of \reef{solnxd3} into $f$ is
\beqa 
f(v_1,v_2,e_1)&\equiv&\frac{1}{16\pi G_{10}}\int dx^H\sqrt{-g}\cal
L\cr &&\cr
&=&\frac{V_3V_5h^2r^3}{16\pi
G_{10}}v_1^{5/2}v_2^{5/2}\left(\frac{20(v_1-v_2)}{v_1v_2h^2}+\frac{h^6}{2v_1^5r^6}e_1^2\right)\,,
\eeqa
where $V_3$ and $V_5$ are the volumes of 3 and 5-sphere with radius one. The electric 
charge carries by the brane is given by 
\beqa
q_1&=&\frac{\prt f}{\prt e_1}=\frac{V_3V_5}{16\pi G_{10}}Q\,.
\eeqa
Now we define the entropy function by taking the Legendre transform of the above 
integral with respect to electric field $e_1$, that is
\beqa 
F(v_1,v_2,q_1)&\equiv&e_1\frac{\prt f}{\prt e_1}-f\cr &&\cr
&=&\frac{V_3V_5h^2r^3}{16\pi
G_{10}}v_1^{5/2}v_2^{5/2}\left(-\frac{20(v_1-v_2)}{v_1v_2h^2}+\frac{h^6}{2v_1^5r^6}e_1^2\right)\,.
\eeqa
Substituting the value of $e_1$ and solving the equations of motion 
\beqa
\frac{\prt F}{\prt v_i}\,=\,0\,, && i=1,2\labell{eom1}\,,
\eeqa 
one finds the following solution
\beqa  
v_1\,=\,1\,,\,\, v_2\,=\,1\,.
\eeqa
Let us now consider the behavior of the entropy function around the above critical point. To 
this end consider the following matrix 
\beq
M_{ij}=\partial_{v_i}\partial_{v_j} F(v_1,v_2)\,.\labell{M}
\eeq
Ignoring the overall constant factor, the eigenvalues of this matrix are $10(5\pm\sqrt{89})$. This 
shows that the critical point $v_1=v_2=1$ is a saddle point of the entropy function. 

Let us now return to the entropy associated with this solution. It is straightforward to find the 
entropy from the Wald formula \cite{Wald}
\beqa 
S_{BH}&=&-\frac{8\pi}{16\pi G_{10}}\int dx^H\sqrt{
g^H}\frac{\prt {\cal L}}{\prt R_{trtr}}g_{tt}g_{rr}\,.\labell{wald10}
\eeqa 
For this background we have $R_{trtr}=\frac{r^4-3r_0^4}{v_1h^2r^4}g_{tt}g_{rr}$ 
and $\sqrt{-g}=v_1\sqrt{g^H}$.
These simplify the entropy relation to 
\beqa
S_{BH}=-\frac{8\pi h^2r^4}{16\pi
G_{10}(r^4-3r_0^4)}\int dx^H\sqrt{-g}\frac{\prt
{\cal L}}{\prt R_{trtr}}R_{trtr}=-\frac{2\pi h^2 r^4}{r^4-3r_0^4}\frac{\prt f_{\l}}{\prt
\l}\bigg|_{\l=1}\,,\labell{Waldlambda}
\eeqa 
where $f_{\l}(v_1,v_2,e_1)$ is an expression similar to $f(v_1,v_2,e_1)$
except that each $R_{trtr}$ Riemann tensor component is scaled by a factor of $\l$.

To find $\frac{\prt f_{\l}}{\prt\l}|_{\l=1}$ using the prescription given 
in \cite{Sen} and \cite{Ghodsi:2006cd}, 
we note that in addition to $R_{trtr}$ the other Riemann tensor 
components $R_{ti_1ti_1},\, R_{ri_1ri_1}$, and
$R_{i_1i_2i_1i_2}$ where $i_1,i_2=1,2,3$ are all proportional to $v_1$, \ie
\beqa
R_{trtr}&=&v_1\frac{3r_0^4-r^4}{h^2r^4}\,,\quad\quad
R_{ri_1ri_1}=v_1\frac{r^4+r_0^4}{h^2(r^4-r_0^4)}\,,\cr &&\cr
R_{ti_1ti_1}&=&v_1\frac{r_0^8-r^8}{h^6r^4}\,,\quad\quad
R_{i_1i_2i_1i_2}=v_1\frac{r^4-r_0^4}{h^6}\,.\labell{4R}
\eeqa
Hence, one should rescale them too. We use the following scaling for these components  
\beq
R_{ti_1ti_1}\rightarrow\l_1 R_{ti_1ti_1}\,,\,\,\,\, 
R_{ri_1ri_1}\rightarrow\l_2 R_{ri_1ri_1}\,,\,\,\,\, 
R_{i_1i_2i_1i_2}\rightarrow\l_3 R_{i_1i_2i_1i_2}\,.\labell{3label} 
\eeq
Now we see that $f_{\l}(v_1,v_2,e_1)$ must be of the form $v_1^{5/2}g(v_2,\l
v_1,e_1v_1^{-5/2},\l_1v_1,\l_2v_1,\l_3v_1)$ for some function $g$.
Then one can show that the following relation holds for $f_\l$ and 
its derivatives with respect to scales, $\l_i, e_1$ and $v_1$:
\beqa
\l\frac{\prt f_{\l}}{\prt \l}+3\l_1\frac{\prt
f_{\l}}{\prt \l_1}+3\l_2\frac{\prt f_{\l}}{\prt
\l_2}+3\l_3\frac{\prt f_{\l}}{\prt \l_3}+\frac{5}{2}e_1\frac{\prt
f_{\l}}{\prt e_1}+v_1\frac{\prt f_{\l}}{\prt
v_1}-\frac{5}{2}f_{\l}&=&0\,.\labell{xscald3}
\eeqa
In addition, there is another relation between the rescaled Riemann 
tensor components at the supergravity level which can be found using \reef{4R}
\beqa
3\frac{\prt f_{\l}}{\prt \l_1}|_{\l_1=1}+3\frac{\prt f_{\l}}{\prt
\l_2}|_{\l_2=1}+3\frac{\prt f_{\l}}{\prt
\l_3}|_{\l_3=1}&=&\frac{3(3r^4+r_0^4)}{r^4-3r_0^4}\frac{\prt f_{\l}}{\prt \l}|_{\l=1}\,.
\eeqa
Replacing the above relation into \reef{xscald3} and using the equations of 
motion, one finds that $\frac{\prt
f_{\l}}{\prt \l}|_{\l=1}=-\frac{1}{4}\frac{r^4-3r_0^4}{r^4}F$. It is easy 
to see that the entropy is proportional to the entropy function up to a constant coefficient, \ie
\beqa 
S_{BH}&=&\frac{\pi h^2}{2}F=\frac{V_3V_5h^2r_0^3}{4G_{10}}\,.\labell{sbh}
\eeqa 
One may write the entropy in terms of temperature. The relation between $r_0$ 
and temperature can be read from the metric which is $r_0=\pi h^2T$, so
\beqa
S_{BH}&=&\frac{\pi^2}{2}N^2V_3T^3\,,
\eeqa 
where we have used the relations $V_5=\pi^3$, $h^4=\frac{N\ka_{10}}{2\pi^{5/2}}$, and
$2\ka_{10}^2=16\pi G_{10}$ where $N$ is the number of D3-branes. This is the 
entropy that has been found in \cite{Gubser:1998nz}. Note that for extremal 
case, $r_0=0$, the entropy function is exactly the same as non-extremal case 
however, the value of entropy is zero.

We have seen that the entropy function formalism works very well here despite the fact 
that the horizon is not attractive. To see the latter fact, we note that the only scalar 
field in this theory is constant everywhere, and it does not appear in the 
Lagrangian. Therefore, it is better to check the attractor property by calculation 
of the proper distance of an arbitrary point from the horizon, \ie
\beq
\rho=\int_{r_0}^r \frac{h}{r}(1-\frac{r_0^4}{r^4})^{-\frac12} dr=\frac12 h 
\log\bigg[\frac{r^2}{r_0^2}+\sqrt{\frac{r^4}{r_0^4}-1}\bigg]\,,
\eeq 
the above value is finite for non-extremal case but it is 
infinite for extremal case \ie $r_0\rightarrow 0$. Hence, although 
the attractor mechanism does not work for this non-extremal case, the entropy 
function formalism works and it gives the correct value of the entropy as the saddle point of the entropy function.
\section{Entropy function for non-extremal $M2$-branes}
The near horizon geometry of non-extremal $M2$-branes is described by the 
Schwarzschild $AdS_{4}\times S^{7}$. 
The most general solution consistent with the symmetry of $AdS_{4}\times S^{7}$ is 
\beqa
ds^2\!\!\!\!&=&\!\!\!\!v_1\left[\frac{y^2}{h^2}\left\{-\left(1-(\frac{y_0}{y})^3\right)dt^2
+\sum_{i=1}^{2}(dx^i)^2\right\}
+\frac{h^2}{4y^2}\left(1-(\frac{y_0}{y})^3\right)^{-1}
dy^2\right]
+v_2h^2(d\Omega_{7})^2,\cr &&\cr
F_{ti_1 i_2y}\!\!\!\!&=&\!\!\!\!3\epsilon_{i_1
i_2}\frac{v_1^{2}}{v_2^{7/2}}\frac{y^{2}}{h^{3}}\equiv
\epsilon_{i_1 i_2}e_1\,,\labell{solnxm2}
\eeqa
where we have defined the new variable $y=r^2/h$. In above $v_1$ and $v_2$ are constants. 
The value of entropy function in this case is given by
\beqa 
F=\frac{V_2V_7h^5y^2}{32\pi
G_{11}}v_1^{2}v_2^{7/2}\left(-\frac{42v_1-48v_2}{v_1v_2h^2}+\frac{2h^4}{v_1^4y^4}e_1^2\right)\,,
\eeqa
where $V_2$ and $V_7$ are the volume of 2 and 7-sphere with radius one. Substituting 
the value of $e_1$ and then solving the equations of motion gives $v_1=v_2=1$.
Moreover, the eigenvalues of the matrix \reef{M} in this case are $3(83\pm\sqrt{12937})$. So 
this shows again that the critical point $v_1=v_2=1$ is the saddle point of the entropy function. 

The entropy associated with this background is given by the Wald formula 
\beqa 
S_{BH}&=&-\frac{8\pi}{16\pi G_{11}}\int dx^H\sqrt{
g^H}\frac{\prt {\cal L}}{\prt R_{tyty}}g_{tt}g_{yy}\,.\labell{wald11}
\eeqa 
For the background \reef{solnxm2} we find $R_{tyty}=\frac{4(y^3-y_0^3)}{v_1h^2y^3}
g_{tt}g_{yy}$ and $\sqrt{-g}\,=\,\frac{1}{2}v_1\sqrt{g^H}$ so that the entropy 
can be written as
\beqa 
S_{BH}&=&-\frac{4\pi h^2y^3}{16\pi G_{11}(y^3-y_0^3)}\int dx^H\sqrt{-g}\frac{\prt
{\cal L}}{\prt R_{tyty}}R_{tyty}=-\frac{\pi h^2 y^3}{y^3-y_0^3}\frac{\prt f_{\l}}{\prt
\l}\bigg|_{\l=1}\,.
\eeqa 
where again we have rescaled every factor $R_{tyty}$ by $\lambda$ in $f_{\lambda}$. In addition to $R_{tyty}$,  
there are three other types of Riemann curvature tensors which are proportional to $v_1$. These 
are $R_{ti_1ti_1}, R_{yi_1yi_1}$ and $R_{i_1i_2i_1i_2}$ with $i_1,i_2=1,2$, \ie
\beqa
R_{tyty}&=&v_1\frac{y_0^3-y^3}{h^2y^3}\,,\,\,\,
R_{yi_1yi_1}=v_1\frac{2y^3+y_0^3}{h^2(2y^3-2y_0^3)}\,,\cr &&\cr
R_{ti_1ti_1}&=&-v_1\frac{4y^6-2y^3y_0^3-2y_0^6}{h^6y^2}\,,\,\,\,
R_{i_1i_2i_1i_2}=v_1\frac{4y^4-4yy_0^3}{h^6}\,.\labell{4Rm2}
\eeqa 
Rescaling them with $\l_1,\l_2$ and $\l_3$ as in \reef{3label} and noting 
that $f_{\l}(v_1,v_2,e_1)$ must be of the form $v_1^{2}g(v_2,\l
v_1,e_1v_1^{-2},\l_1v_1,\l_2v_1,\l_3v_1)$ for some function $g$, one finds 
the following relation: 
\beqa
\l\frac{\prt f_{\l}}{\prt \l}+2\l_1\frac{\prt
f_{\l}}{\prt \l_1}+2\l_2\frac{\prt f_{\l}}{\prt
\l_2}+\l_3\frac{\prt f_{\l}}{\prt \l_3}+2 e_1\frac{\prt
f_{\l}}{\prt e_1}+v_1\frac{\prt f_{\l}}{\prt
v_1}-2f_{\l}&=&0\,,\labell{flm2}
\eeqa 
using \reef{4Rm2}, one finds
\beqa
2\frac{\prt f_{\l}}{\prt \l_1}\bigg|_{\l_1=1}+2\frac{\prt f_{\l}}{\prt
\l_2}\bigg|_{\l_2=1}+\frac{\prt f_{\l}}{\prt
\l_3}\bigg|_{\l_3=1}&=&\frac{5y^3+y_0^3}{y^3-y_0^3}\frac{\prt
f_{\l}}{\prt \l}\bigg|_{\l=1}\,.
\eeqa 
Replacing the above relation into the \reef{flm2}, one finds $\frac{\prt f_{\l}}{\prt
\l}|_{\l=1}=-\frac{y^3-y_0^3}{3y^3}F$ and therefore
\beqa 
S_{BH}=\frac{\pi h^2}{3}F=\frac{V_2V_7h^5y_0^2}{4G_{11}}\,,
\eeqa 
this gives a non-zero value for entropy. One may write the entropy in terms 
of temperature. From the metric \reef{solnxm2} we find the relation between 
non-extremality parameter and temperature as $y_0=2\pi h^2T/3$. So the entropy becomes
\beqa
S_{BH}&=&2^{7/2}3^{-3}\pi^2V_2N^{3/2}T^2
\eeqa 
where we have used the relations $V_7=\pi^4/3$, $h^9=N^{3/2}\frac{\ka^2_{11}
\sqrt{2}}{\pi^{5}}$ and $2\ka_{11}^2=16\pi G_{11}$ where $N$ is the number of M2-branes. 
This is the entropy, which has been found in \cite{Gubser:1998nz}. Note again that 
the extremal case can be found by taking $y_0=0$. The result for entropy function 
is exactly the same as non-extremal case but the value of entropy is zero.

To check the attractor mechanism, we note that there is no scalar field in this 
theory so we calculate  the proper distance of an arbitrary point from the horizon, \ie
\beq
\rho=\int_{y_0}^y \frac{h}{2y}(1-\frac{y_0^3}{y^3})^{-\frac12} dy=\frac13 h 
\log\bigg[(\frac{y}{y_0})^{\frac32}+\sqrt{(\frac{y}{y_0})^{3}-1}\bigg]\,,
\eeq 
which is finite for the non-extremal case but is infinite when $y_0\rightarrow 0$ 
in the extremal case. This shows again that although the horizon is not attractive 
point, the entropy function formalism works and it gives the correct value for the entropy as the 
saddle point of the entropy function.

\section{Entropy function of non-extremal $M5$-branes}
For non-extremal M5-branes the background is Schwarzschild $AdS_7\times S^4$ and 
the general solution consistent with this symmetry is 
\beqa
ds^2_{11}&=&v_1\left[\frac{y^2}{h^2}\left\{-\left(1-(\frac{y_0}{y})^6\right)
dt^2 +\sum_{i=1}^{5}(dx^i)^2\right\}
+4\frac{h^2}{y^2}\left(1-(\frac{y_0}{y})^6\right)^{-1}
dy^2\right]\cr &&\cr
&+&v_2h^2(d\Omega_4)^2\,,\quad\quad F_{ti_1\cdots i_5r}=6\epsilon_{i_1\cdots
i_5}\frac{v_1^{7/2}}{v_2^2}\frac{y^5}{h^6}\equiv
\epsilon_{i_1\cdots i_5}e_1\,,\labell{solnxm5}
\eeqa
where we have used the new coordinate $y=\sqrt{hr}$. The entropy function for 
this background is
\beqa 
F &=&\frac{2V_5V_4y^5}{16\pi
G_{11}h}v_1^{7/2}v_2^{2}\left(-\frac{24v_1-21v_2}{2v_1v_2h^2}+\frac{h^{10}}
{8v_1^7y^{10}}e_1^2\right)\,,
\eeqa
where $V_5$ and $V_4$ are the volume of 5 and 4-sphere with radius one.
Substituting the value of $e_1$ and solving the equations of motion results
$v_1=v_2=1$. 
The eigenvalues of the matrix \reef{M} in this case are $\frac38(29\pm\sqrt{12937})$. 
Therefore it shows that the critical point $v_1=v_2=1$ is a 
saddle point of the entropy function.

Let us now turn to the entropy associated with  this solution. The Wald formula 
in \reef{wald11} still holds here.
Using the fact that for this background $R_{tyty}=\frac{y^6-10y_0^6}{4v_1h^2y^6}g_{tt}g_{yy}$ 
and $\sqrt{-g}=2v_1\sqrt{g^H}$
one finds
\beqa 
S_{BH}=-\frac{16\pi h^2y^6}{16\pi
G_{11}(y^6-10y_0^6)}\int dx^H\sqrt{-g}\frac{\prt
{\cal L}}{\prt R_{tyty}}R_{tyty}
=-\frac{4\pi h^2 y^6}{y^6-10y_0^6}\frac{\prt f_{\l}}{\prt
\l}\bigg|_{\l=1}\,,
\eeqa 
where we have rescaled $R_{tyty}$ in $f_{\lambda}$. There are other Riemann tensor components 
proportional to $v_1$. These are  $R_{ti_1ti_1},\, R_{yi_1yi_1}$ and
$R_{i_1i_2i_1i_2}$ with $i_1,i_2=1...5$, \ie
\beqa
R_{tyty}&=&v_1\frac{10y_0^6-y^6}{h^2y^6}\,,\,\,\,
R_{yi_1yi_1}=v_1\frac{y^6+2y_0^6}{h^2(y^6-y_0^6)}\,,\cr &&\cr
R_{ti_1ti_1}&=&-v_1\frac{y^{12}+y^6y_0^6-2y_0^{12}}{4h^6y^8}\,,\,\,\,
R_{i_1i_2i_1i_2}=v_1\frac{y^6-y_0^6}{4h^6y^2}\,.
\eeqa 
We rescale them by $\l_1,\l_2$ and $\l_3$. Nothing that
$f_{\l}(v_1,v_2,e_1)$ must be of the general form $v_1^{7/2}g(v_2,\l
v_1,e_1v_1^{-7/2},\l_1v_1,\l_2v_1,\l_3v_1)$ for some function $g$, one finds
\beqa 
\l\frac{\prt f_{\l}}{\prt \l}+5\l_1\frac{\prt
f_{\l}}{\prt \l_1}+5\l_2\frac{\prt f_{\l}}{\prt
\l_2}+10\l_3\frac{\prt f_{\l}}{\prt \l_3}+\frac{7}{2}e_1\frac{\prt
f_{\l}}{\prt e_1}+v_1\frac{\prt f_{\l}}{\prt
v_1}-\frac{7}{2}f_{\l}&=&0\,,\labell{flm5}
\eeqa 
One finds also the following relation at the supergravity level:
\beqa
5\frac{\prt f_{\l}}{\prt \l_1}\bigg|_{\l_1=1}+5\frac{\prt f_{\l}}{\prt
\l_2}\bigg|_{\l_2=1}+10\frac{\prt f_{\l}}{\prt
\l_3}\bigg|_{\l_3=1}&=&10\frac{2y^6+y_0^6}{y^6-10y_0^6}\frac{\prt f_{\l}}{\prt
\l}\bigg|_{\l=1}\,.
\eeqa 
Replacing the above relation in \reef{flm5} one can show that $\frac{\prt f_{\l}}
{\prt \l}|_{\l=1}=-\frac{1}{6}\frac{y^6-10y_0^6}{y^6}F$ and therefore
\beqa 
S_{BH}=\frac{2\pi h^2}{3}F=\frac{V_5V_4y_0^5}{4G_{11}h}\,,
\eeqa 
this gives non-zero result. To write the entropy in terms of temperature we 
use $y_0=\frac{4\pi h^2T}{3}$ then
\beqa
S_{BH}&=&2^73^{-6}\pi^3N^3V_5T^5\,,
\eeqa 
where we have used the relations $V_4=\frac{8\pi^2}{3}$, $h^9=\frac{N^3\ka_{11}^2}
{2^7\pi^5}$ and $2\ka_{11}^2=16\pi G_{11}$ where $N$ is the number of M5-branes. This is 
in agreement with the result in \cite{Gubser:1998nz}.

We look now to the attractor mechanism. As we see, there is no scalar field in this 
case so we check the attractor property by calculation of the proper distance of 
an arbitrary point from the horizon
\beq
\rho=\int_{y_0}^y \frac{2h}{y}(1-\frac{y_0^6}{y^6})^{-\frac12} dy=\frac23 h 
\log\bigg[(\frac{y}{y_0})^{3}+\sqrt{(\frac{y}{y_0})^{6}-1}\,\bigg]\,,
\eeq 
which is finite for the non-extremal case but is infinite when $y_0\rightarrow 0$ 
in the extremal case. This shows again that although the horizon is not attractive 
for the non-extremal case, the entropy function formalism works and it gives the 
correct value for the entropy as the saddle point of the entropy function.

\section{Higher derivative terms for non-extremal $D3$-branes}
In the previous sections, we have seen that the entropy function works at two 
derivatives level. It will be interesting to consider stringy effects and 
look at the entropy function mechanism again. To this end, we consider 
the higher derivative corrections coming from string theory. To next leading 
order the Lagrangian of IIB theory in Einstein frame is given by
\beqa
S&=&\frac{1}{16\pi G_{10}}\int
d^{10}x\,\sqrt{-g}\bigg\{
R-\frac{1}{2}g^{\mu\nu}\prt_{\mu}\phi\prt_{\nu}\phi-\frac{1}{2}\sum\frac{1}
{n!}F_{(n)}^2+\gamma
e^{-3\phi/2}W\bigg\}\,,\labell{eff1}
\eeqa 
where $\gamma=\frac18\zeta(3)(\al')^3$ and $W$ can be written in terms of 
the Weyl tensors
\beq
W=C^{hmnk}C_{pmnq}{C_h}^{rsp}{C^q}_{rsk}+\frac12 C^{hkmn}C_{pqmn}{C_h}^{rsp}{C^q}_{rsk}\,.
\eeq
In what follows, we will show that \reef{solnxd3} is no longer a solution of the above action.
We calculate the contribution of the above higher
derivative terms to the entropy function $F$
\beqa 
\delta F\!\!\!\!&=&\!\!\!\!-\frac{\gamma}{16\pi G_{10}}\int dx^H\sqrt{-g}W= \cr &&\cr
\!\!\!\!&=&\!\!\!\!-\gamma\frac{V_3V_5}{16\pi G_{10}}\frac{180}{h^6r^{13}(v_1v_2)^{3/2}}
\left[\frac{35}{1944}(v_1-v_2)^4r^{16}+
\frac{1}{6}v_2^2(v_1-v_2)^2r_0^8r^8+v_2^4r_0^{16}\right]\!\!,
\eeqa
By variation of  $F+\delta F$ with respect to $ v_1$ and $v_2$ one
finds the equations of motion. Since these equations are valid only up
to first order of $\gamma$, we consider the following perturbative
solutions
\beqa 
v_1=1+\gamma x\,,\,\,\, v_2=1+\gamma y\,.
\eeqa 
For extremal case, $r_0=0$, the corrections are zero, \ie $v_1=1=v_2$. Again the value 
of entropy is proportional to $r^3$ which gives zero. This is due to the fact that 
$AdS_5\times S^5$ is an exact solution.

For non-extremal case, by replacing the above solutions into the equations of motion, one
finds the following relations 
\beqa
\frac{\prt(F+\delta F)}{\prt v_1}=0\,\longrightarrow
&&3x+5y=\frac{27}{h^6}\left(\frac{r_0}{r}\right)^{16}\,,\cr &&\cr
\frac{\prt(F+\delta F)}{\prt v_2}=0\,\longrightarrow
&&5x-13y=-\frac{45}{h^6}\left(\frac{r_0}{r}\right)^{16}\,,
\eeqa
these equations are consistent, and give the following result
\beqa 
v_1=1+\frac{63}{32h^6}\left(\frac{r_0}{r}\right)^{16}\gamma\,&,&\,
v_2=1+\frac{135}{32h^6}\left(\frac{r_0}{r}\right)^{16}\gamma\,.
\eeqa
However, they are functions of $r$. This is inconsistent with our assumption that 
$v_1$ and $v_2$ are constants!. 
So it seems that the deformed geometry \reef{solnxd3} is not the solution of equations 
of motion when we consider higher derivative terms. 
Hence the entropy function formalism does not work when higher derivative corrections are added to the 
effective action. The same thing happens for $M2$ 
and $M5$-branes. 

This is related to the fact that in the presence of the higher derivative terms the 
solution is not the Schwarzschild $AdS$ anymore. The ansatz for the metric should 
be \cite{Gubser:1998nz}
\beqa
ds^2=r^2(-e^{2a+8b}dt^2+e^{2b}dr^2+d\vec{x}^2)+e^{2c}d\Omega_5^2\,,\labell{anz}
\eeqa
where $a, b$ and $c$ are functions of $r$ and we have chosen $h=1$. The solution for these 
functions at linear order of $\gamma$ gives a metric which is not
the Schwarzschild $AdS$ \cite{Gubser:1998nz}.
Using the ansatz \reef{anz}, one realizes that the horizon area does not modify so 
the entropy is given by \reef{Waldlambda} where now $f_\l$ is replaced by $f_\l+f_\l^W$, \ie
\beqa
S_{BH}=-\frac{2\pi h^2 r^4}{r^4-3r_0^4}\frac{\prt (f_{\l}+f_\l^W)}{\prt
\l}\bigg|_{\l=1}\,,\labell{Waldlambdac}
\eeqa  
where the function $f^W$ is given by
\beq
f^W=\frac{\gamma}{16\pi G_{10}}\int dx^H \sqrt{-g} e^{-\frac32 \phi}W\,.
\eeq
The first term in \reef{Waldlambdac} give the same result as before, \ie \reef{sbh}. 
The second term is proportional to $\gamma$, so to the first order of $\gamma$ one 
has to replace the Schwarzschild $AdS$ solution \reef{solnxd3} in $\frac{\partial f_\l^W}
{\partial\l}$ which gives 
\beq
\frac{\partial f_\l^W}{\partial\l}\bigg|_{\l=1}=-120\frac{V_3V_5}{16\pi G_{10}}
\frac{(r^4-3r_0^4)r_0^{12}}{ r^{13}}\,.
\eeq
Finally the entropy will be
\beq
S_{BH}=\frac{V_3V_5 h^2}{4 G_{10}}r_0^3\bigg(1+60\gamma+{\mathcal{O}}(\gamma^2)\bigg)\,.
\eeq
In terms of temperature  \cite{Gubser:1998nz}, $T=\frac{r_0}{\pi}(1+15\gamma)$, one finds
\beq
S_{BH}=\frac{\pi^2}{2}N^2V_3T^3\bigg(1+15\gamma\bigg)\,.
\eeq
This is the entropy that has been found in \cite{Gubser:1998nz} using the free energy formalism.

\section{Discussion}
In this paper, we have studied in details the entropy function formalism for 
non-extremal $D3$, $M2$ and $M5$-branes. We have shown that the entropy function 
can be applied to find the entropy of these solutions at tree level. The entropy 
function in all cases has a saddle point and the entropy is given by the value 
of this function at this point.

We have studied  non-extremal black branes, which have either no moduli or constant 
moduli. The
non-extremal black holes which have non-constant moduli, has been studied 
in \cite{Cai:2007ik}. One may expect
that in this case also the entropy function should have a saddle point. To see this more explicitly 
let us consider the 5 dimensional non-extremal black holes in $IIB$ theory 
compactified on $T^4\times S^1$ with the following $BTZ\times S^2$ near 
horizon geometry \cite{Cai:2007ik}:  
\begin{eqnarray}
ds^{2}&=&v_{1}\bigg[-\frac{(\rho^{2}-\rho_{+}^{2})(\rho^{2}-\rho_{-}^{2})}{\rho^{2}}dt^2
+\frac{4\rho^{2}}{(\rho^{2}-\rho_{+}^{2})(\rho^{2}-\rho_{-}^{2})}d\rho^{2}
 \nonumber \\
      &+&{\rho^{2}}(dz-\frac{\rho_{+}\rho_{-}}{\rho^{2}}dt)^2\bigg]
+v_{2}d\Omega_{2}^{2}, \nonumber \\
e^{-2\phi}&=&u_{s},~~~e^{2\psi}=u_{T},~~~e^{\frac{\psi_{1}}{2}}=u_{1},
\nonumber
\\ F^{(5)}_{tz\rho}&=&e_{1}=\frac{\rho
u_{1}}{u_{T}}\frac{v_{1}^{\frac{3}{2}}}{v_{2}},~~~
H^{(5)}_{\theta\varphi}=-\frac{1}{2}\sin\theta,~~~G_{\theta\varphi}=-\frac{1}{2}\sin\theta,
\end{eqnarray}
where $e^{2\psi}$ and $e^{\frac{\psi_{1}}{2}}$ denote the single
moduli for $T^{4}$ and $S^{1}$ respectively. We refer the reader 
to \cite{Cai:2007ik} for details.
The entropy function in this case is proportional to
\begin{eqnarray}
F&\sim &v_{1}^{\frac{3}{2}}v_{2}u_{T}u_{1}
 \left [u_{s}(\frac{3v_{2}-4v_{1}}{2v_{1}v_{2}}+\frac{1}{2u_{1}^{2}v_{2}^{2}})
   +\frac{1}{2v_{2}^{2}}+\frac{e_{1}^{2}}{2u_{1}^{2}
  \rho^{2}v_{1}^{3}} \right ].
\end{eqnarray}
The solution to the equations of motion 
\begin{equation}
\frac{\partial F}{\partial u_{i}}=0,~~i=s,T,1,~~~~\frac{\partial
F}{\partial v_{j}}=0,~~j=1,2,
\end{equation}
is $v_1=v_2=v,\, u_{s}=\frac{1}{v},\,  u_{T}=1,\, u_{1}=\frac{1}{\sqrt{v}}$.
As can be seen, these equations of motion cannot fix all the moduli so 
one expects that the entropy function has a flat direction \cite{Sen}. 
To study the behavior of the entropy function around the above critical 
point, consider the following matrix: 
\beq
M_{ij}=\partial_{\phi_i}\partial_{\phi_j}F\,,\quad \phi_i=\{v_1,v_2,u_s, u_T, u_1\}\,. 
\eeq
The eigenvalues of this matrix for $v=1$ are
\beq
(4.81,\, -3.34,\, 2.23,\, 0.55,\, 0)\,.\labell{evch}
\eeq
The negative eigenvalue indicates that the critical point is a saddle point. 
Moreover as anticipated above one of the eigenvalues is zero. 

We have seen in sections 3, 4 and 5 that the entropy function has one minimum 
and one maximum in the 
directions specified by the sizes of $AdS_{p+2}$ and $S^{D-(p+2)}$. This might be 
related to the fact that curvature of $AdS_{p+2}$ is negative and the curvature 
of $S^{D-(p+2)}$ is positive. This property should be independent of the 
attractiveness of the black holes. So one expects that this property holds even for 
extremal solutions with $AdS_{p+2}\times S^{D-(p+2)}$ near horizon.
To see this, consider the Dyonic black holes in Heterotic string theory 
compactified on $M\times S^1 \times \tilde{S}^1$ where $M$ is a four dimensional 
compact manifold and $S^1$ and $\tilde{S}^1$ are circles \cite{Sen}. The near 
horizon geometry is given by
\beqa
ds^2&=&v_1\bigg(-r^2 dt^2+\frac{dr^2}{r^2}\bigg)+v_2\bigg(d\theta^2+\sin^2\theta 
d\phi^2\bigg)\,,\cr &&\cr
S&=&u_S\,,\quad R=u_R\,, \quad \tilde{R}=u_{\tilde{R}}\,, \cr &&\cr
F_{rt}^{(1)}&=&e_1\,, \quad F_{rt}^{(3)}=e_3\,, \quad F_{\theta\phi}^{(2)}=p_2\,, 
\quad F_{\theta\phi}^{(4)}=p_4\,,
\eeqa  
where $S$ is dilaton and $R$ and $\tilde{R}$ are radii of the circles.
The entropy function in terms of the electric and magnetic charges $q_1,q_3,p_2,p_4$ 
is given by
\beq
F=\frac{\pi}{4} v_1v_2u_S\bigg[\frac{2}{v_1}-\frac{2}{v_2}+\frac{8q_1^2}{u_R^2u_S^2v_2^2}
+\frac{8u_R^2q_3^2}
{u_S^2v_2^2}+\frac{2u_{\tilde{R}}^2\,p_2^2}{16\pi^2v_2^2}+\frac{2p_4^2}
{16\pi^2u_{\tilde{R}}^{2}\,v_2^2}\bigg]+4\pi u_S\,,
\eeq 
where the charge quantization gives $q_1=\frac{n}{2}, q_3=\frac{w}{2}, p_2=4\pi\tilde{k}, 
p_4=4\pi\tilde{w}$.
Solving equations of motion gives rise to the following solutions for scalars
\beqa
v_1=v_2=4\tilde{n}\tilde{w}+8\,,\quad
u_S=\sqrt{\frac{n w}{\tilde{n}\tilde{w}+4}}\,,\quad u_R=\sqrt{\frac{n}{w}}\,,
\quad u_{\tilde{R}}=\sqrt{\frac{\tilde{w}}{\tilde{n}}}\,. 
\eeqa 
We can construct the following matrix as before:
\beq
M_{ij}=\partial_{\phi_i}\partial_{\phi_j}F\,,\quad \phi_i=\{v_1,v_2,u_S, u_R, u_{\tilde{R}}\}\,, 
\eeq
For the case that $n=w=\tilde{n}=\tilde{w}=1$ the eigenvalues are 
\beq
(70.44,\, 28.10,\, 5.62,\, -0.11,\, 0.03)\,.\labell{evsen}
\eeq
We see that as expected the critical point is a saddle point. 

The eigenvalues \reef{evch} and \reef{evsen} indicate that the critical point 
in both non-extremal and extremal solutions are the saddle points of the entropy 
function. However, the attractiveness of the solutions cannot be seen from these eigenvalues.
The attractiveness can be studied either by the proper distance of an arbitrary 
point from the horizon \cite{Cai:2007ik} or by looking at the effective potential 
for the moduli fields.
The effective potential can be read from the entropy function by inserting in 
the values of sizes $v_1$ and $v_2$.
Doing this one finds that the eigenvalues of the matrix $M_{ij}$ constructed 
from the effective potential, have negative values in non-extremal case whereas 
for extremal case, all the eigenvalues are positive.    

The entropy function formalism works for those black holes/branes that their 
near horizon is an extension of $AdS$ space. 
The near horizon (throat approximation) of the p-brane solutions ($D3,\,M2,\,M5$-branes) 
that we have studied are the Schwarzschild $AdS$ times sphere. For other p-branes 
this near horizon is not a product space so the entropy function formalism 
does not work. One may consider instead the near horizon (not 
the throat approximation) of the non-extremal p-brane solutions which is a 
product of the Rindler space times a sphere. It can easily be checked that 
the entropy function formalism does not work for this
space. \footnote{It has been shown in \cite{Astefanesei:2006sy} that the 
attractor mechanism does not work for the Rindler space.}

We have seen in the section 6 that the higher derivative corrections modify 
the tree level solutions such that the near horizon (throat approximation) 
is not the Schwarzschild $AdS$ anymore.
Consequently, the entropy function formalism does not work for these cases. 
Hence, we have used the Wald formula to find the value of entropy directly. 
It would be interesting to find a non-extremal solution where the higher 
derivative corrections respect the symmetries of the tree level solution \ie $AdS$. 
In those cases, one would expect to find the entropy function including the higher 
derivative corrections by using the entropy function formalism \cite{wip}.  


\end{document}